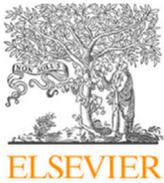
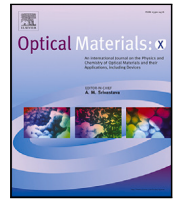

Invited article

# Tuning exciton recombination rates in doped transition metal dichalcogenides

Theresa Kuechle [a,1], Sebastian Klimmer [a,1], Margarita Lapteva [a], Tarlan Hamzayev [a], Antony George [b,c], Andrey Turchanin [b,c], Torsten Fritz [a,c], Carsten Ronning [a,c], Marco Gruenewald [a], Giancarlo Soavi [a,c,*]

[a] *Institute of Solid State Physics, Friedrich Schiller University Jena, Helmholtzweg 5, Jena, 07743, Germany*
[b] *Institute of Physical Chemistry, Friedrich Schiller University Jena, Lessingstraße 10, Jena, 07743, Germany*
[c] *Abbe Center of Photonics, Friedrich Schiller University Jena, Albert-Einstein-Str. 6, Jena, 07745, Germany*

A B S T R A C T

Monolayer transition metal dichalcogenides (TMDs) are direct gap semiconductors that hold great promise for advanced applications in photonics and optoelectronics. Understanding the interplay between their radiative and non-radiative recombination pathways is thus of crucial importance not only for fundamental studies but also for the design of future nanoscale on-chip devices. Here, we investigate the interplay between doping and exciton–exciton annihilation (EEA) and their impact on the photoluminescence quantum yield in different TMD samples and related heterostructures. We demonstrate that the EEA threshold increases in highly doped samples, where the radiative and non-radiative recombination of trions dominates.

## 1. Introduction

More than a decade ago, the experimental evidence of the crossover from an indirect to a direct gap obtained by exfoliating semiconducting transition metal dichalcogenides (TMDs) from bulk (3D) to single-layer (2D) [1,2] has ignited an enormous interest in the scientific community. Besides the fundamental studies on the photophysics of excitons and related quasi-particles (charged excitons/trions, biexcitons, charged biexcitons/quintons *etc.*) in ultra-clean 2D quantum confined systems [3], TMDs also provide new exciting solutions for optoelectronic [4], valleytronic [5] and nonlinear optical [6,7] applications. In particular, the giant enhancement of the photoluminescence (PL) quantum yield (QY) in the monolayer limit [1] holds great promise for integrated, flexible and high-speed light emitting devices [8]. In this regard, it is paramount to consider that light emission in TMDs strongly depends on intrinsic and extrinsic factors such as doping, environment, defects, strain and the photoexcited carrier density [3,9].

This work focuses on the interplay between doping and photoexcited carrier concentration in monolayer TMDs and explores their impact on the radiative and non-radiative recombination pathways of excitons and trions. On the one hand, doping in TMDs is responsible of trion (*i.e.*, charged exciton) formation [10,11]. Since trions' PL emission occurs at lower energies with respect to neutral zero-momentum excitons, tuning of the doping in TMDs offers a viable tool for active devices with tuneable optical properties at the cost of a reduced PL-QY [12]. The reduced PL-QY at large doping values is a direct consequence of the ultrafast non-radiative trion recombination time [13] that dominates over the radiative recombination channels [14]. Indeed, undoped TMDs display a close to unity PL-QY, in contrast to the <1% PL-QY of doped samples [13,15]. On the other hand, the radiative and non-radiative recombination pathways in TMDs are strongly affected by the excited state carrier concentration. For instance, due to their reduced dimensionality and in analogy with other quantum confined systems such as quantum dots [16], graphene nanoribbons [17] and carbon nanotubes [18], TMDs display strong exciton–exciton annihilation (EEA), a nonlinear process where one exciton annihilates (*i.e.*, recombines to the ground state) and transfers its energy to another exciton that is promoted to a higher energy level [19,20]. In this regard, EEA represents another source of non-radiative recombination that limits the QY of light emitting devices based on TMDs [21].

Understanding the interplay between doping and EEA in TMDs is thus of crucial importance for the design of advanced nanoscale light emitting devices. In this work we show that in highly doped TMDs trion radiative and non-radiative recombination dominates over the nonlinear EEA, leading to a linear scaling of the PL power dependent emission, also at high values of the photoexcited carrier density. On the other hand, EEA plays a crucial role in samples with low doping. Our results can be interpreted with a rate equation model that takes into account all the radiative and non-radiative recombination pathways of excitons and trions in TMDs as a function of doping (*i.e.*, trion concentration) and generation rate (*i.e.*, photoexcited carrier concentration). This work sheds light on the complex excited photophysics of TMDs and on the effect of doping and nonlinear recombination pathways on their radiative emission.

## 2. Experimental

### 2.1. Samples fabrication and PL characterization

The influence of the doping on the PL-QY was first investigated by chemical functionalization of a $MoS_2$ monolayer. To this end, we employed a high quality monolayer $MoS_2$ fabricated with a modified chemical vapor deposition (CVD) method [22,23] on thermally oxidized silicon/silicon dioxide ($Si/SiO_2$) substrates. After initial optical studies, the same $MoS_2$ flake was chemically doped with potassium (K-$MoS_2$) using an alkali metal dispenser (SAES Getters). Chemical functionalization of $MoS_2$ with K leads to an increase of the electron concentration, namely n-doping [24]. In order to confirm the effect of K functionalization, in Fig. 1 we show the normalized PL spectra for pristine $MoS_2$ and K-$MoS_2$. Both PL spectra were simultaneously fitted with two Lorentzian functions for the A exciton and trion [25]. For the B exciton, the best fit was obtained using a Gaussian function, possibly due to contributions from B trions [26]. From the fitting, we obtained that neutral excitons lie at ∼1.86 eV (A exciton) and ∼2.01 eV (B exciton), in good agreement with literature [1,2,27,28]. The $A^-$ trion peak appears at ∼30 meV below the A neutral exciton, thus displaying a binding energy in agreement with previously reported theoretical [29,30] and experimental values [12,31]. In particular, we note that in the K-$MoS_2$ sample the amplitude of the trion peak is much higher compared to the pristine $MoS_2$ sample, confirming that K chemical functionalization leads to increased n-doping [24]. In addition, we prepared a layered heterostructure composed of a CVD-grown $MoS_2$ monolayer and exfoliated few-layer graphene (FLG/$MoS_2$). This layered heterostructure was fabricated using the polymethylmethacrylate (PMMA)-assisted wet transfer method [32]. Finally, we studied and optically characterized a layered heterostructure composed of $MoSe_2$ and few-layer graphene (FLG/$MoSe_2$). Both materials were exfoliated from commercial bulk crystals (HQ Graphene) and the heterostructure was fabricated using a deterministic dry-transfer method [33]. Compared to $MoS_2$, $MoSe_2$ has a lower inherent doping [13] and thus offers a suitable platform to study neutral exciton PL emission and EEA.

### 2.2. Experimental setup

All the measurements were performed at room temperature with a custom built optical microscope setup, as sketched in Fig. 2(a). We performed PL experiments using a continuous wave helium–neon laser (LGK 7786 P100, RPMC Lasers, Inc.) at a wavelength of 543 nm (∼ 2.28 eV). We tuned the power with a continuously variable neutral density filter (NDC-50C-4M, Thorlabs) and we subsequently focused the beam on the sample with a 50x objective, thus obtaining a spot diameter at the focus position of (1.6±0.1) μm. Additionally, a collimated white light source (WL) and a CMOS (C-B5, OPTIKA) were integrated in the setup to image the sample and therefore further characterize it via optical contrast. The back-scattered PL signal was spectrally filtered with a dichroic mirror (DM, DMSP567, Thorlabs) and a longpass filter (84-745, Edmund Optics) and finally detected with a monochromator (Acton Research SpectraPro-150, 300 lines per mm) equipped with a Si CCD (Roper Scientific, SpectrumMM 250B with UV-enhancement coating).

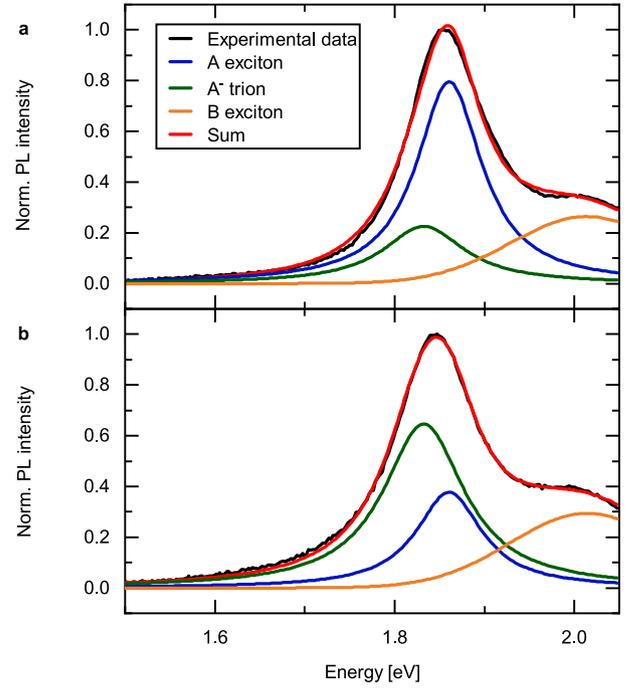

**Fig. 1. PL spectra of $MoS_2$ and functionalized K-$MoS_2$ samples.** (a)–(b) Normalized PL spectra with fits for different exciton contributions. Central energy and width of individual species have been kept identical for comparison. (a) Normalized PL spectrum for $MoS_2$. The PL spectrum is dominated by the neutral A exciton peak. (b) Normalized PL spectrum for K-$MoS_2$. The K functionalization produces a strong n-doping, as confirmed by the enhanced trion emission.

## 3. Results and discussion

### 3.1. Modeling of the PL power dependent emission

We focus on the impact of doping on the radiative and non-radiative exciton recombination dynamics. This can be investigated by looking at the power dependence in steady-state PL experiments. In order to understand the impact of doping and EEA on the steady-state PL, we briefly introduce a simple rate equation model that takes into account all the possible radiative and non-radiative recombination channels for A excitons and trions [34]. First, the time-evolution of the quasiparticle density consisting of excitons and trions in all our samples can be described by the equation [13]:

$$\frac{d(n_X + n_T)}{dt} = G - \frac{n_X}{\tau_X} - \frac{n_T}{\tau_T} - \gamma n_X^2 \qquad (1)$$

where $G$ is the exciton generation rate which is proportional to the initial excitation density and therefore the light intensity $I$. Further, $n_X$ and $n_T$ are the neutral A exciton and negative trion densities, respectively, $\tau_X$ and $\tau_T$ are the associated lifetimes, which are again composed of radiative (r) and non-radiative (nr) contributions: $\frac{1}{\tau_X} \equiv \frac{1}{\tau_X^r} + \frac{1}{\tau_X^{nr}}$, $\frac{1}{\tau_T} \equiv \frac{1}{\tau_T^r} + \frac{1}{\tau_T^{nr}}$. Last, $\gamma$ is the EEA coefficient [15,35,36]. All these radiative and non-radiative recombination channels are schematically depicted in Fig. 3(a–d). In the steady-state condition, when $\frac{d(n_X + n_T)}{dt} = 0$, we have:

$$G = \frac{n_X}{\tau_X^r} + \frac{n_T}{\tau_T^r} + \frac{n_T}{\tau_T^{nr}} + \gamma n_X^2 \qquad (2)$$

Note that Eq. (2) does not include the non-radiative neutral exciton lifetime, since we are assuming that neutral exciton recombination is entirely radiative in a monolayer [13]. Furthermore, since our experiments are conducted at room temperature, the influence of dark excitons [37] can also be neglected, as dark to bright transitions are thermally activated. In the following, we will further neglect the non-radiative trion recombination time (∼ tens of ps), which is orders of magnitude faster than the radiative recombination time (∼ 100 ns) but it does not affect the power scaling of the PL signal $PL = \frac{n_X}{\tau_X^r} + \frac{n_T}{\tau_T^r}$.

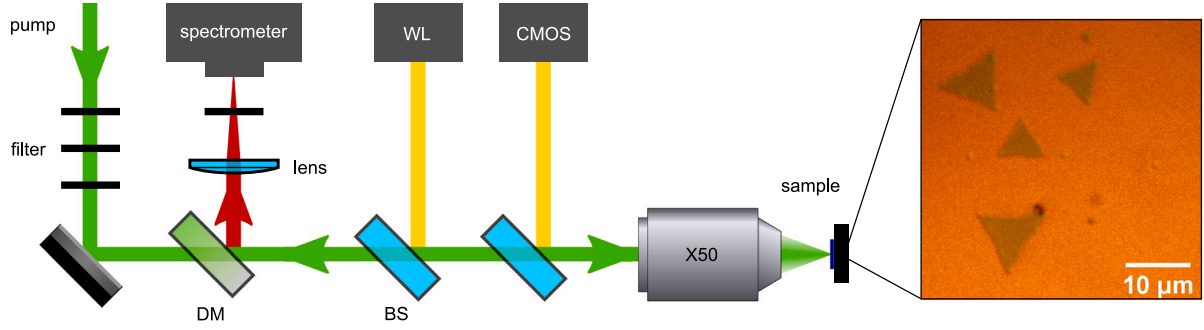

**Fig. 2. Sketch of the experimental setup.** The pump laser beam (543 nm) is spectrally purified by a laser line filter and attenuated using neutral density filters. The beam is subsequently focused onto the MoS$_2$ sample (CVD-grown) with a 50x microscope objective, which is imaged with a collimated white light source (WL) and a camera (CMOS) that can be coupled in with beamsplitters (BS). The back-scattered PL signal is separated with a dichroic mirror (DM), further spectrally filtered and subsequently detected with a spectrometer. The green line represents the excitation beam and the red line represents the PL signal. (For interpretation of the references to color in this figure legend, the reader is referred to the web version of this article.)

From here, we can analyze different doping and excitation density conditions. In an intrinsic system, we expect a negligible trion density ($n_T \approx 0$). Thus we can neglect contribution from radiative recombination of trions and re-write Eq. (2) as:

$$G = \frac{n_X}{\tau_X^r} + \gamma n_X^2 \quad (3)$$

Based on the generation rate, we can again distinguish between two different conditions. At low excitation density, the influence of EEA is negligible ($\frac{n_X}{\tau_X^r} \gg \gamma n_X^2$). In this case, the steady-state PL signal is given by:

$$PL = \frac{n_X}{\tau_X^r} = G \propto I, \quad (4)$$

corresponding to a linear scaling of the emitted PL signal with respect to the incident excitation intensity. On the contrary, at a high exciton density, EEA dominates ($\gamma n_X^2 \gg \frac{n_X}{\tau_X^r}$) [13] and the PL signal becomes:

$$PL = \frac{n_X}{\tau_X^r} = \frac{\sqrt{G\gamma^{-1}}}{\tau_X^r} \propto \sqrt{I}. \quad (5)$$

Next, we discuss the PL signal in a strongly doped system. Here, we can derive the trion density in the steady-state condition from the mass action law $n_T = T n_X n_e$, where $T$ denotes the trion formation coefficient [38] and $n_e$ is the density of free electrons. We further define the total negative charge density $N = n_T + n_e$ composed by negative trions and free electrons. Subsequently, we can write the trion density as a function of the neutral exciton density as [13]:

$$n_T = \frac{T n_X}{1 + T n_X} N. \quad (6)$$

When we plug Eq. (6) in Eq. (2) we obtain:

$$G = \frac{n_X}{\tau_X^r} + \frac{T n_X}{1 + T n_X} \frac{N}{\tau_T^r} + \gamma n_X^2 \quad (7)$$

At low excitation density, and therefore low generation rate, we have a low exciton density ($1 \gg T n_X$) and the trion and exciton densities are proportional to each other ($n_T \approx T n_X N$). In parallel, in this condition, EEA is again negligible and we thus obtain the simplified expression for the generation rate:

$$G = n_X \left( \frac{1}{\tau_X^r} + \frac{TN}{\tau_T^r} \right). \quad (8)$$

In analogy with the undoped case, we can re-write the expected PL signal as:

$$PL = \frac{n_X}{\tau_X^r} + \frac{n_T}{\tau_T^r} = n_X \left( \frac{1}{\tau_X^r} + \frac{TN}{\tau_T^r} \right) = G \propto I. \quad (9)$$

Finally, we consider the combination of a strongly doped system and a high excitation density. As already mentioned, the contribution of EEA increases for a higher excitation intensity and therefore a higher generation rate. Moreover, in this condition, the trion density approaches the total negative charge density ($n_T \approx N$) and thus modifies the generation rate:

$$G = \frac{n_X}{\tau_X^r} + \frac{N}{\tau_T^r} + \gamma n_X^2. \quad (10)$$

We note that depending on the generation rate either EEA or radiative trion recombination can dominate the recombination dynamics depending on $\gamma$ and $T$ [13].

### 3.2. Experimental results for PL power dependent emission

We performed power-dependent PL measurements on the pristine MoS$_2$, K-MoS$_2$ and FLG/MoS$_2$. Exemplary PL spectra for pristine MoS$_2$ at different values of the incident intensity are shown in Fig. 4(a) while Fig. 4(b) shows a comparison between the PL power dependence on MoS$_2$ (black line) and K-MoS$_2$ (red line). The PL power dependence is always evaluated from the total number of counts per second in the energy interval ±100 meV around the A emission peak, since the ratio of the different contributing exciton species does not change with increasing excitation density. Further, the data sets were subsequently normalized for ease of comparison. Both samples show a slope in log–log scale close to one, indicating that EEA is negligible. In particular, the slope of K-MoS$_2$ (∼ 1.0) is higher compared to the slope of pristine MoS$_2$ (∼ 0.93). Although sample inhomogeneity, e.g., due to strain or high defect density, has a strong impact on the linear and nonlinear optical properties of TMDs [39,40], we do not expect large point-to-point deviations in our high quality CVD-grown materials [22,23].

We compared also the results obtained from pristine MoS$_2$ and K-MoS$_2$ with the FLG/MoS$_2$ heterostructure. Here, following optical excitation, a large fraction of the photoexcited carriers in MoS$_2$ are transferred to FLG on an ultrafast timescale (∼ 1 ps) [41,42]. This effectively corresponds to a reduction of the doping and, as a consequence, to a change in the PL power dependence from linear (slope of 1) to sublinear (slope of 0.5), as expected in a system where EEA dominates. Indeed, our experiments show a slope of 0.87, indicating a regime where both effects (doping and EEA) coexist and play a role, i.e., a crossover between low and high excitation density within the framework of our model. It is worth noting that in our experiments, we used an incident intensity in the range 0.5 kWcm$^{-2}$ to 16.5 kWcm$^{-2}$ corresponding to a generation rate ranging from $1.4 \times 10^{20}$ cm$^{-2}$s$^{-1}$ to $45.1 \times 10^{20}$ cm$^{-2}$s$^{-1}$ considering an absorption of $\alpha \approx 0.1$ [43] and assuming that every absorbed photon generates one exciton. According to Ref. [13], at this generation rate the radiative trion recombination and EEA are comparable in MoS$_2$ for typical values of the intrinsic doping.

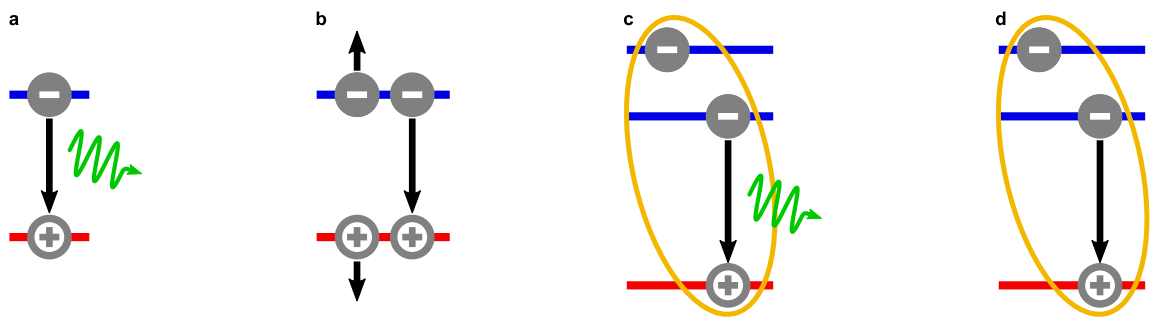

**Fig. 3. Exciton and trion recombination pathways.** (a)–(d) Radiative and non-radiative recombination processes contributing to Eq. (1). (a) Radiative exciton recombination, (b) exciton–exciton annihilation, (c) radiative trion recombination and (d) non-radiative trion recombination.

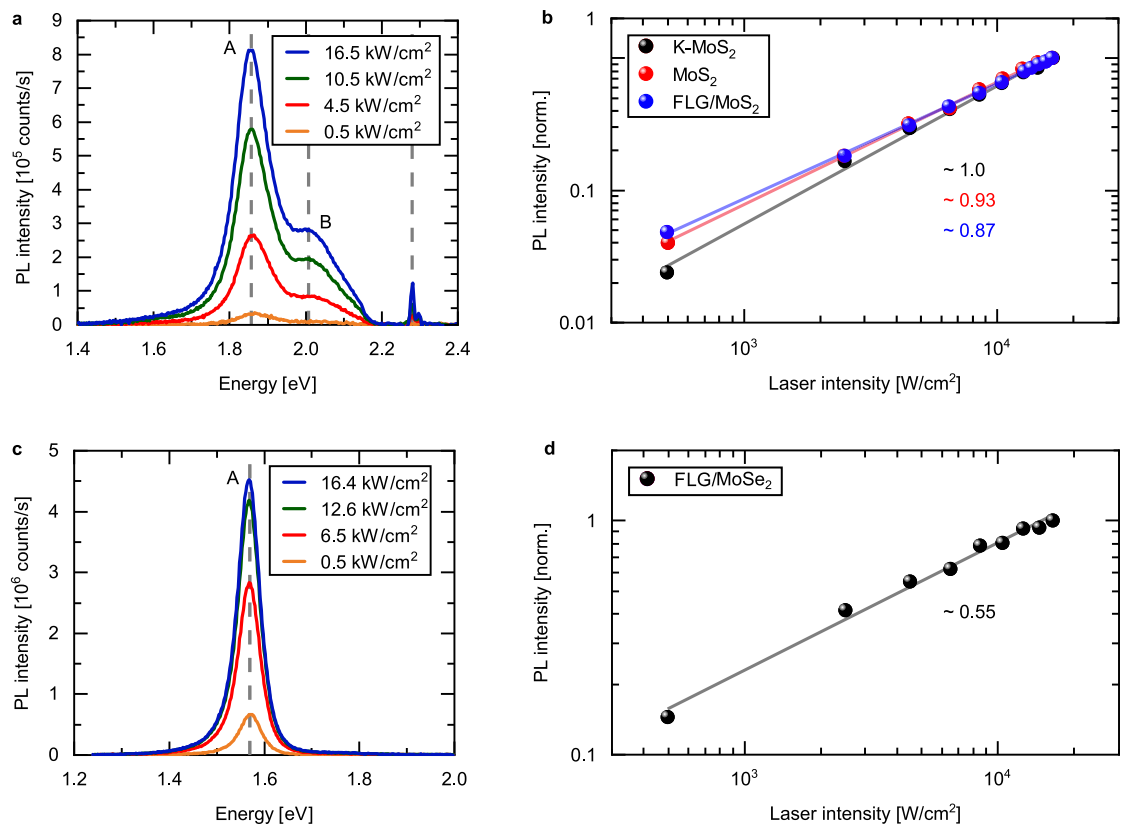

**Fig. 4. PL power dependence.** (a) PL spectra of $MoS_2$ for different values of the excitation intensity. Two spectral features at 1.86 eV and 2.01 eV are assigned to the A and B excitons, respectively. The weak peak at 2.28 eV is a residual from the excitation laser. (b) Normalized PL power dependence in log–log scale showing the total detected counts in the energy interval ±100 meV around the A emission peak for $MoS_2$, K-$MoS_2$ and FLG/$MoS_2$. (c) PL spectra of FLG/$MoSe_2$ for different values of the excitation intensity. The spectral feature at 1.57 eV is assigned to the A exciton, whereas no B exciton peak is visible. (d) Normalized PL power dependence in log–log scale showing the total detected counts in the energy interval ±100 meV around the A emission peak for FLG/$MoSe_2$. (For interpretation of the references to color in this figure legend, the reader is referred to the web version of this article.)

To further confirm our hypothesis, we repeated the PL power dependent experiments on the FLG/$MoSe_2$ sample (Fig. 4(c)). $MoSe_2$ is typically nearly intrinsic [13] and, in addition, the FLG removes any residual free carriers due to charge transfer, as recently demonstrated in the case of four different tungsten- and molybdenum-based FLG/TMD heterostructures [44] and as we also observed for FLG/$MoS_2$. We note that in layered FLG/TMD heterostructures, two types of charge transfer are simultaneously present. On the one hand, a static charge transfer (without any optical excitation) shifts the TMD Fermi level/doping [44]. On the other hand, ultrafast charge transfer following optical excitation can lead to a reduction of the photoexcited carrier density [41,45]. The former will affect the term $\frac{n_T}{\tau_T}$ in Eq. (1), while the latter only shifts the EEA threshold and is neglected in our simplified model. Here, we identify one prominent feature in the spectrum at 1.57 eV, which we assign to the A exciton, in agreement with previous studies [46,47].

Moreover, we do not observe any emission from the B exciton, confirming the high quality of our sample [27]. The power dependence at the A exciton shows a slope close to 0.5 in a double-logarithmic plot (Fig. 4(d)). According to our model, this is a clear indication of a regime where doping is negligible and the PL emission is dominated by EEA, as expected for this values of excitation intensities in naturally undoped selenium-based TMDs [13].

## 4. Conclusions

In conclusion, we have studied the effect of doping and exciton–exciton annihilation on the power dependent radiative emission of different TMD samples. In particular, we were able to increase the n-doping in pristine $MoS_2$ by chemical functionalization with potassium and to reduce it by fabrication of a layered heterostructure composed of $MoS_2$ and few-layer graphene. In addition, we studied an almost un-doped heterostructure composed of $MoSe_2$ and few-layer graphene. For the samples with high doping, PL power dependent measurements show a trend that is close to linear. On the contrary, in undoped samples, the radiative emission scales with the square root of the incident power. These results are fully captured by a simple rate equation model that shows how the trion radiative and non-radiative emission dominates in the case of high doping even for very large values of the excitation fluence (or generation rate) while EEA plays a crucial role already at low fluences in undoped samples.

## CRediT authorship contribution statement

**Theresa Kuechle:** Conceived the experiments, Performed the PL measurements, Wrote the manuscript, Discussion and commented on the manuscript. **Sebastian Klimmer:** Wrote the manuscript, Discussion and commented on the manuscript. **Margarita Lapteva:** Fabricated and provided the high quality TMD monolayer and heterostructure samples, Discussion and commented on the manuscript. **Tarlan Hamzayev:** Fabricated and provided the high quality TMD monolayer and heterostructure samples, Discussion and commented on the manuscript. **Antony George:** Fabricated and provided the high quality TMD monolayer and heterostructure samples, Discussion and commented on the manuscript. **Andrey Turchanin:** Fabricated and provided the high quality TMD monolayer and heterostructure samples, Discussion and commented on the manuscript. **Torsten Fritz:** Discussion and commented on the manuscript. **Carsten Ronning:** Conceived the experiments, Discussion and commented on the manuscript. **Marco Gruenewald:** Conceived the experiments, Performed the PL measurements, Discussion and commented on the manuscript. **Giancarlo Soavi:** Conceived the experiments, Wrote the manuscript, Discussion and commented on the manuscript.

## Declaration of competing interest

The authors declare that they have no known competing financial interests or personal relationships that could have appeared to influence the work reported in this paper.

## Acknowledgments

This work was supported by the European Union's Horizon 2020 Research and Innovation program under Grant Agreement GrapheneCore3 881603 (G.S.). The authors acknowledge the German Research Foundation DFG (CRC 1375 NOA project numbers B2 (A.T.), B5 (G.S.) and C5 (C.R.)) and the Daimler und Benz foundation, Germany for financial support (G.S.).